\documentclass[12pt]{iopart}

\usepackage{epsfig,array,amssymb,psfrag,graphicx,tensor,color,mathrsfs}

\definecolor{blue2}{RGB}{0, 30, 255}
\definecolor{dblue}{RGB}{0, 0, 150}

\usepackage[colorlinks=true, linkcolor=blue2, citecolor=blue2, urlcolor=dblue]{hyperref}

\newcommand{\beq}{\begin{equation}}
\newcommand{\eeq}{\end{equation}}
\newcommand{\bea}{\begin{eqnarray}}
\newcommand{\eea}{\end{eqnarray}}

\newcommand{\RR}{\mathbb{R}}

\newcommand{\mc}[1]{\mathcal{#1}}
\newcommand{\mb}[1]{\mathbf{#1}}

\newcommand{\ii}{\mathrm{i}}
\newcommand{\ee}{\mathrm{e}}

\newcommand{\cL}{\mathscr{L}}

\newcommand{\intd}{\mathrm{d}}

\makeatletter \@addtoreset{equation}{section} \makeatother 
\renewcommand{\theequation}{\arabic{section}.\arabic{equation}}

\begin{document}

\title{Noether currents for the Teukolsky Master Equation}

\author{G\'abor Zsolt T\'oth} 
\address{Institute for Particle and Nuclear Physics, 
Wigner Research Centre for Physics,
Konkoly-Thege Mikl\'os \'ut 29-33,
1121 Budapest, Hungary}
\ead{toth.gabor.zsolt@wigner.mta.hu}

\begin{abstract}
Conserved currents associated with the time translation and axial symmetries of the Kerr spacetime
and with scaling symmetry
are constructed for the Teukolsky Master Equation (TME). Three partly different approaches are taken, 
of which the third one applies only to the spacetime symmetries. 
The results yielded by the three approaches,
which correspond to three variants of Noether's theorem,
are essentially the same, nevertheless. The construction includes the embedding of the TME into a larger system of 
equations, which admits a Lagrangian and turns out to consist of two TMEs with opposite spin weight. 
The currents thus involve two independent solutions of the TME with opposite spin weights.
The first approach provides an example of the application of an extension of Noether's theorem to nonvariational 
differential equations. 
This extension is also reviewed in general form.
The variant of Noether's theorem applied in the third approach is a generalization of the 
standard construction of conserved currents associated with spacetime symmetries in general relativity, in which 
the currents are obtained by the contraction of the symmetric energy-momentum tensor with the relevant Killing vector fields.
Symmetries and conserved currents related to boundary conditions are introduced as well, 
and Noether's theorem and its variant for nonvariational differential equations are extended to them. 
The extension of the latter variant is used to construct conserved currents related to the Sommerfeld boundary condition. 
\end{abstract}

\noindent{\it Keywords\/}: Kerr spacetime, Teukolsky master equation, conserved current, Noether's theorem, boundary condition

\maketitle





\section{Introduction}
\label{sec.intr}

Conservation laws are important properties of dynamical systems, as they allow one to make statements about the dynamics 
without solving the equations of motion. 
They also tend to be conspicuous features, because their validity extends to all orbits of the system.
The conservation laws associated with spacetime symmetries and internal symmetries are among the most important and characteristic ones.

The aim of the present paper is to construct conserved currents associated with time translation, axial rotation and scaling symmetry 
for the Teukolsky Master Equation (TME) \cite{teukolsky1,teukolsky2,s32a,s32b}, which 
is a wave equation that governs the evolution of the extreme spin weight Newman--Penrose components \cite{NP,GHP} in Kinnersley tetrad
of the Maxwell, the linearized gravitational, spin-$1/2$ (neutrino) or spin-$3/2$ fields in Kerr spacetime, 
and plays an important role in the analysis of these fields.

We look for conserved currents for the TME mainly because they can be used to verify numerical solutions of the TME generated by computer (see \cite{tme1}-\cite{GHBN} for numerical studies of the solutions of the TME).
Since the codes used for such numerical simulations are complicated, it is important to test them, 
and one way of doing this is to check that the numerically generated solutions indeed satisfy the 
conservation laws relevant for them. Examples of this usage of conserved currents can be found in \cite{FFGR}-\cite{CLR}.
A further motivation for looking for conserved currents for the TME is provided by the recent interest in the symmetries 
and associated currents
of the Maxwell and linearized gravitational fields in Kerr spacetime \cite{AA}-\cite{AB}. 
An important objective of the latter studies is to find currents which can be used in obtaining decay estimates for these fields. 
Whether the currents found in this paper are useful in this context is not obvious, however, 
since the currents used for obtaining decay estimates are usually required to have suitable positivity properties.

In numerical computations the time evolution is often calculated only in a spatially limited domain, 
and suitable conditions are imposed on the fields at the boundaries to ensure that the time evolution is well defined. 
In addition to verifying that the numerically generated solutions satisfy the equations of motion,
one might thus also be interested in testing whether the intended boundary conditions are also satisfied.
A secondary aim of the paper is to address this problem, although it is less important in practice, 
since the boundary conditions are usually much simpler than the equations of motion.
It should also be noted that if the location of a boundary is chosen so that the future light cones are directed
outward from the computational domain at the boundary, then it is not necessary to impose any boundary condition 
(see \cite{ZT,HBB,RT} for examples of numerical studies in which boundaries of this type are chosen).

We construct the currents associated with time translation and axial rotation symmetry in three partially different ways, 
by applying three variants of Noether's theorem. 
The results yielded by these three approaches are essentially the same, nevertheless. 
Since the TME does not follow from a Lagrangian, 
in the first approach a relatively less known variant of Noether's theorem is applied, 
which is valid for any differential equation, regardless of whether it is an Euler--Lagrange equation corresponding to a Lagrangian 
or not.
It involves the embedding of the TME into a larger system, which has a Lagrangian. 
This system turns out to consist of a pair of TMEs with opposite spin weight. 
The constructed currents thus involve two independent solutions of the TME with opposite spin weights.    
In the second approach the second order Lagrangian obtained in the first approach is replaced by a first order one by adding
total divergences, and then the standard Noether construction is applied to get the conserved currents. 
In the third approach a further version of Noether's theorem is applied, which
makes use of the fact that the Lagrangian obtained in the second (or first) approach is diffeomorphism invariant in a certain sense, 
and that the time translations and the rotations are special diffeomorphisms. 
This version of Noether's theorem is a generalization of the standard construction in general relativity in which the conserved currents associated with spacetime symmetries are obtained by contracting the energy-momentum tensor with the Killing vector fields that generate the symmetries. The standard construction cannot be applied because of the presence of a fixed vector field and a fixed scalar field,
apart from the source term, in the TME.
For the scaling symmetry only the first two approaches will be considered, since the third one is not applicable.

For dealing with the problem of the verification of boundary conditions, 
we extend the first two variants of Noether's theorem to boundary conditions and their symmetries in a general setting.
This involves introducing definitions for the symmetries of boundary conditions, conserved currents at boundaries, 
and Lagrangians for boundary conditions. 
Then the variant of the extended Noether theorem that is suitable for boundary conditions that do not follow from a Lagrangian
is applied to the Sommerfeld boundary condition, which is often used in numerical computations of solutions of wave equations (see \cite{tme2}, for example).

Before discussing the particular case of the TME in Section \ref{sec.teuk}, we review briefly in general form 
the variant of the Noether construction that pertains to arbitrary differential equations in Section \ref{sec.sde}.
The standard Noether construction is reviewed in \ref{sec.nthr}.
These short reviews are included for the sake of completeness and because we believe that they can be helpful for readers who 
intend to find further conserved currents for the TME or for other differential equations. 
Section \ref{sec.sde} also contains a simplification in comparison with the literature
and allows anticommuting fields.
For a detailed account of the last variant of Noether's theorem mentioned above, the reader is referred to \cite{T}. 

The general discussion of boundary conditions and their symmetries and conserved currents can be found in Section \ref{sec.bcurrents} 
and in \ref{sec.bcurr}. These parts of Section \ref{sec.sde} and \ref{sec.nthr} are new, to our knowledge.  
The conserved currents related to the Sommerfeld boundary condition in the case of the TME are discussed in Section \ref{sec.tmebc}.

\section{Noether currents for symmetries of differential equations}
\label{sec.sde}

In this section, it is discussed how conserved currents can be constructed for symmetries of systems of differential equations.
In the first step, the differential equations are embedded into a larger set
of equations which are the Euler--Lagrange equations corresponding to a suitable Lagrangian density function, and then the Noether construction described in \ref{sec.nthr} is applied in a particular way to obtain conserved currents associated with the symmetries of the original system of differential equations. Further details on this and closely related constructions can be found in \cite{BCA}-\cite{Ibr2}. In Section \ref{sec.bcurrents}, the construction is extended to boundary conditions and their symmetries. 

In the following, $\Phi_i$ is a collection of fields or field components indexed by the general index $i$, 
$M$ denotes the base manifold in which the fields propagate, $x^\mu$, $\mu=0,1,\dots, D$, 
are coordinates that cover an open domain $U$ in $M$, $D+1$ is the dimension of $M$.
$\Phi_i$ can be real or complex valued, 
and they are also allowed to be anticommuting (Grassmann algebra valued) for some values of $i$. 
For derivatives with respect to anticommuting variables, the following sign convention will be used: 
if $\theta$ is an anticommuting variable and $E$ is an expression of the form 
$E_1\theta E_2$, then $\frac{\partial E}{\partial \theta}= (-1)^n E_1 E_2$, where $n=0$ if $E_2$ is even and $n=1$ if 
$E_2$ is odd. The square bracket notation $F[\phi]$, where $\phi_i$ are some fields indexed by $i$,
will be used to indicate that $F$ is a local function of $\phi_i$, 
which means that it is a function of $x^\mu$, $\phi_i(x^\mu)$ and finitely many derivatives of $\phi_i(x^\mu)$.

Let us consider a system of differential equations
\beq
\label{eq.de}
F^a(x^\mu,\Phi_i(x^\mu),\partial_\nu\Phi_i(x^\mu),\partial_{\nu\lambda}\Phi_i(x^\mu),\dots)=0
\eeq
for $\Phi_i$. 
The index $a$ labeling the equations is generally not related to the index $i$ that labels the fields, and 
$F^a$ are assumed to have definite commutation properties, i.e.\ they are either even or odd. 
It is also assumed that $F^a$ is differentiable as many times as necessary, but further assumptions on $F^a$ 
(e.g.\ nondegeneracy) are not made, unless explicitly stated.

In order to embed (\ref{eq.de}) into a system of Euler--Lagrange equations, one extends first the set of fields by  
adding a set of auxiliary fields $\rho_a$, which have the same commutation properties as $F^a$, 
and then one takes the Lagrangian density function
\beq
\label{eq.l}
L[\Phi,\rho]=F^a[\Phi]\rho_a\, .
\eeq
As the index of $\rho_a$ indicates, there is one auxiliary field corresponding to each equation in the system (\ref{eq.de}). 
The Euler--Lagrange equations following from (\ref{eq.l}) for $\rho_a$ are just (\ref{eq.de}), and the Euler--Lagrange equations for $\Phi_i$, 
\beq
\hspace{-2.2cm}\mb{E}[\Phi,\rho]^i =
\frac{\delta L}{\delta \Phi_i} 
=\frac{\partial (F^a\rho_a)}{\partial \Phi_i}-\partial_\mu\frac{\partial (F^a\rho_a)}{\partial (\partial_\mu\Phi_i)}
+\partial_{\mu\nu}\frac{\partial (F^a\rho_a)}{\partial (\partial_{\mu\nu}\Phi_i)}-
\partial_{\mu\nu\lambda}\frac{\partial (F^a\rho_a)}{\partial (\partial_{\mu\nu\lambda}\Phi_i)}+
\dots = 0\, ,
\label{eq.elaux}
\eeq
constitute a further
set of equations, which are linear in $\rho_a$. 
The complete set of Euler--Lagrange equations are satisfied if $\Phi_i$ satisfy (\ref{eq.de}) and $\rho_a=0$, therefore
the Lagrangian system defined by (\ref{eq.l}) indeed properly contains (\ref{eq.de}).
If (\ref{eq.de}) are linear equations, 
then (\ref{eq.elaux})  does not depend on $\Phi_i$ and their derivatives.
Furthermore, (\ref{eq.elaux}) is the adjoint of (\ref{eq.de}) in this case. Generally, (\ref{eq.elaux}) is the adjoint of the 
linearization of (\ref{eq.de}) (see \cite{BCA, Anco2,Anco1, Wald} for further details on adjoint equations).
The above idea for embedding the system (\ref{eq.de}) into a Lagrangian system appears, for example, 
in \cite{Olver,Anco1,Ibr1,Ibr2}.

After embedding (\ref{eq.de}) into the Lagrangian system specified by (\ref{eq.l}), one can try to find symmetries of $L$, 
and then one can construct the associated conserved currents according to the prescription in \ref{sec.nthr}.
In particular, if (\ref{eq.de}) has a symmetry, then $L$ also has a corresponding symmetry, as described below.

A transformation $\Phi_i\to\Phi_i+\varsigma\,\delta\Phi_i$ is a called a continuous symmetry of (\ref{eq.de}), if 
\beq
\label{eq.sc}
\delta F^a=\frac{dF^a[\Phi+\varsigma\,\delta\Phi]}{d\varsigma}|_{\varsigma=0}=0
\eeq
holds for any solution of (\ref{eq.de}). This symmetry condition is the infinitesimal form of the requirement that a symmetry is a transformation that maps a solution of (\ref{eq.de}) into another solution. One can also consider partial symmetries, 
which are characterized by the condition that (\ref{eq.sc}) holds only for a subset of all solutions of (\ref{eq.de}). 

If (\ref{eq.de}) is linear and $O$ is a not necessarily linear symmetry operator for (\ref{eq.de}), 
i.e.\ a mapping on the space of the field configurations that maps solutions of (\ref{eq.de}) into solutions,
then the transformation characterized by $\delta\Phi_i= (O\Phi)_i$ is obviously a continuous symmetry of (\ref{eq.de}).

If $\Phi_i\to\Phi_i+\varsigma\,\delta\Phi_i$ is a symmetry of (\ref{eq.de}), then
$\delta L=F^a\delta \rho_a+\delta F^a \rho_a$ is clearly zero if $F^a=0$, for any choice of $\delta\rho_a$.
This means that $\Phi_i\to\Phi_i+\varsigma\,\delta\Phi_i$, $\rho_a\to\rho_a+\varsigma\,\delta\rho_a$ is also a symmetry
of $L$ with $K^\mu=0$ in the sense defined in \ref{sec.nthr}, with arbitrary $\delta\rho_a$.
Since $K^\mu=0$, the associated Noether current is $j^\mu$ (see (\ref{eq.elc}) for the definition of $j^\mu$). More explicitly,
\bea
j^\mu[\rho,\Phi,\delta\Phi] & = &  \frac{\partial (F^a\rho_a)}{\partial (\partial_\mu\Phi_i)}\delta \Phi_i+
\Biggl(\frac{\partial (F^a\rho_a)}{\partial (\partial_{\mu\nu}\Phi_i)}\partial_\nu \delta \Phi_i
-\partial_\nu\frac{\partial (F^a\rho_a)}{\partial (\partial_{\mu\nu}\Phi_i)} \delta \Phi_i\Biggr)\nonumber\\
&& \hspace{-14mm} + \Biggl(\frac{\partial (F^a\rho_a)}{\partial (\partial_{\mu\nu\lambda}\Phi_i)}\partial_{\nu\lambda}\delta\Phi_i
-\partial_\nu\frac{\partial (F^a\rho_a)}{\partial (\partial_{\mu\nu\lambda}\Phi_i)}\partial_\lambda\delta\Phi_i
+\partial_{\nu\lambda}\frac{\partial (F^a\rho_a)}{\partial (\partial_{\mu\nu\lambda}\Phi_i)}\delta\Phi_i\Biggr)+\dots\, . \nonumber \\
\label{eq.elc2}
\eea
$j^\mu$ is conserved if $\Phi_i$ satisfy (\ref{eq.de}) and $\rho_a$ satisfy the auxiliary equations (\ref{eq.elaux}). 
Since $L$ does not depend on the derivatives of $\rho_a$, 
$j^\mu$ does not depend on the choice of $\delta\rho_a$.  
$j^\mu$ is linear in $\rho_a$, therefore it is necessary to find nonzero solutions of (\ref{eq.elaux}) for $\rho_a$ in order
to obtain nonzero $j^\mu$.
The foregoing arguments apply to partial symmetries as well, 
with the obvious modification that the conservation of $j^\mu$ follows only for those solutions of (\ref{eq.de}) for which 
the symmetry condition (\ref{eq.sc}) holds.

A remarkable feature of the above construction is that $K^\mu=0$ can be chosen in the application of Noether's standard theorem, 
i.e.\ it is not necessary to search for a suitable $K^\mu$, 
and the ambiguity of the conserved current associated with the choice of $K^\mu$ is avoided. 
We also note that in the application of Noether's theorem in the above argument 
we used the symmetry condition (\ref{eq.K}) only on-shell,
and this simplified the argument significantly, as we did not need to think about the off-shell values of $\delta L$, 
which depend also on $\delta \rho_a$. In Section 2.2 of \cite{Anco1} and in \cite{Ibr1}, the authors had to find suitable values for 
$\delta \rho_a$, as they considered the symmetry condition on $L$ also off-shell.
A disadvantage of the construction is that it is necessary to solve also (\ref{eq.elaux})
for $\rho_a$ in order to obtain actual conserved currents. 
On the other hand, if it is possible to find many solutions of (\ref{eq.elaux}) for any solution of (\ref{eq.de}),
then the construction yields many conserved currents for each symmetry of (\ref{eq.de}).

If (\ref{eq.de}) is a possibly inhomogeneous linear system of equations, 
then $\frac{\partial (F^a\rho_a)}{\partial (\partial_{\dots}\Phi_i)}$ do not depend on $\Phi_i$ and their derivatives, 
therefore $j^\mu$ becomes a local bilinear current $\hat{\mc{S}}^\mu[\rho,\delta\Phi]=j^\mu[\rho,\Phi,\delta\Phi]$.
If $\rho_a\to\rho_a+\varsigma\,\delta\rho_a$ is a symmetry of the auxiliary equations (\ref{eq.elaux}), then 
$\delta\rho_a$ is also a solution of these equations, therefore
$\hat{\mc{S}}^\mu[\delta\rho,\delta\Phi]$ is also a conserved current.
If (\ref{eq.de}) is homogeneous linear, then the scaling transformation under which $\delta\Phi_i=\Phi_i$ is a symmetry of (\ref{eq.de}),
thus $\hat{\mc{S}}^\mu[\rho,\Phi]$ is a conserved current. 
In the literature (see e.g.\ \cite{Wald}), $\hat{\mc{S}}^\mu[\rho,\Phi]$ is often obtained in a direct way without referring to scaling symmetry and Noether's theorem: if $\mc{D}^{ai}$ are the differential operators that specify $F^a$, i.e.\ $F^a[\Phi]=\mc{D}^{ai}\Phi_i$, 
then by applying the product rule of derivatives one finds 
\beq
\label{eq.adj}
(\mc{D}^{ai}\Phi_i) \rho_a = \partial_\mu \hat{\mc{S}}^\mu[\rho,\Phi] + (\mc{D}^{Tia}\rho_a) \Phi_i\, ,
\eeq
where $\mc{D}^{Tia}$
are differential operators
and $\mc{D}^{Tia}\rho_a$ are the expressions $\frac{\delta L}{ \delta \Phi_i}$ appearing in the auxiliary (adjoint) system (\ref{eq.elaux}). From (\ref{eq.adj}) it is obvious that $\hat{\mc{S}}^\mu[\rho,\Phi]$ is a conserved current.

\subsection{Conserved currents associated with boundary conditions and their symmetries}
\label{sec.bcurrents}

In \ref{sec.bcurr} the extension of Noether's theorem to boundary conditions that follow from a Lagrangian is discussed.
In this section the embedding approach described above is extended to boundary conditions, allowing one to find 
conserved currents for boundary conditions regardless of whether they follow from a Lagrangian or not. 

Let $\mc{B}$ be a boundary of $M$ and
\beq
\label{eq.bcF}
F^b_\mc{B}(x^\mu,\Phi_i(x^\mu),\partial_\nu\Phi_i(x^\mu),\partial_{\nu\lambda}\Phi_i(x^\mu),\dots)|_{\mc{B}}=0
\eeq
a system of boundary conditions at $\mc{B}$ indexed by some index $b$. One introduces then the auxiliary fields
$\rho_{\mc{B}b}$ and the auxiliary Lagrangian density function
\beq
\label{eq.bLagr}
L_{\mc{B}}[\Phi,\rho_\mc{B}]= F^b_\mc{B}[\Phi]\rho_{\mc{B}b}\, .
\eeq
$\rho_{\mc{B}b}$ and $F^b_\mc{B}$ are assumed to be defined in a neighbouhood of $\mc{B}$.
The boundary Euler--Lagrange equations $\frac{\delta L_\mc{B}}{\delta \rho_{\mc{B}b}}|_{\mc{B}} = 0$
are the boundary conditions (\ref{eq.bcF}), 
whereas the boundary Euler--Lagrange equations $\frac{\delta L_\mc{B}}{\delta \Phi_i }|_{\mc{B}} = 0$ constitute a set of auxiliary equations that are linear in $\rho_{\mc{B}b}$.

A transformation $\Phi_i\to\Phi_i+\varsigma\,\delta\Phi_i$ is a called a symmetry of (\ref{eq.bcF}), if 
\beq
\label{eq.Bsc}
\delta F^b_\mc{B}|_\mc{B}=
\Biggl(\frac{dF^b_\mc{B}[\Phi+\varsigma\,\delta\Phi]}{d\varsigma}|_{\varsigma=0}\Biggr)|_\mc{B}=0
\eeq
holds whenever (\ref{eq.bcF}) is satisfied. 
It can be seen in the same way as in the case of (\ref{eq.de}) that  
a symmetry of (\ref{eq.bcF}) is also a symmetry of $L_{\mc{B}}$ with $K_{\mc{B}}^\mu = 0$, 
therefore a Noether current can be constructed according to the prescription of \ref{sec.bcurr}.  
Since $K_{\mc{B}}^\mu = 0$, 
the formula that gives this current is (\ref{eq.elc2}) with the obvious modifications $F^a\to F^b_{\mc{B}}$, $\rho_a\to \rho_{\mc{B}b}$.

For an example, let us consider the Neumann boundary condition $(\partial_1 \Phi)|_{x^1=0}=0$ at $x^1=0$ for a single scalar field.
The Lagrangian (\ref{eq.bLagr}) is $\partial_1 \Phi \rho_{\mc{B}}$ and the auxiliary equation for $\rho_{\mc{B}}$
is $-(\partial_1\rho_{\mc{B}})|_{x^1=0} = 0$. The current $j_{\mc{B}}^\mu$ is 
$\frac{\partial L_{\mc{B}}}{\partial(\partial_\mu\Phi)}\delta\Phi = \delta^\mu_1 \rho_{\mc{B}}\delta\Phi$. 
The Neumann boundary condition has translation symmetry in any direction $h^\mu$, where $h^\mu$ is a constant vector. 
$\delta\Phi=-h^\mu\partial_\mu \Phi$ under translations, thus the corresponding boundary conserved currents are
$J^\mu_{\mc{B}}=-\delta^\mu_1 \rho_{\mc{B}}h^\nu\partial_\nu\Phi$. The Neumann boundary condition also has scaling symmetry, 
under which $\delta\Phi=\Phi$, and the corresponding boundary conserved current is 
$J^\mu_{\mc{B}} = \delta^\mu_1 \rho_{\mc{B}}\Phi$.

We note that the embedding method does not give any conserved current (i.e.\ it gives zero) for the Dirichlet boundary condition, 
since the associated Lagrangian (see \ref{sec.bcurr}) does not depend on any derivatives of the fields.

\section{Conserved currents for the Teukolsky Master Equation}
\label{sec.teuk}

Let us recall that the TME can be written in the following compact form, found in \cite{BCJR}:
\beq
\label{eq.tme}
[(\nabla^\mu + s\Gamma^\mu)(\nabla_\mu +s\Gamma_\mu)-4s^2\Psi_2]\psi^{(s)}=4\pi T^{(s)}\, .
\eeq
The line element of the Kerr metric, which is the background metric in (\ref{eq.tme}), 
reads 
\beq
\label{eq.metric}
ds^2=\left(1-\frac{2Mr}{\Sigma}\right)dt^2+\frac{4 a r M\sin^2\theta}{\Sigma}dt d\phi
-\frac{\Sigma}{\Delta}dr^2 - \Sigma d\theta^2 -\frac{\Gamma }{\Sigma}\sin^2\theta d\phi^2
\eeq 
in Boyer-Lindquist coordinates $(t,r,\theta,\phi)$, with
$\Sigma=r^2+a^2\cos^2\theta$, $\Delta=r^2-2Mr+a^2$, $\Gamma=(r^2+a^2)^2-a^2 \Delta\sin^2\theta$. 
$M$ denotes the mass of the black hole
and $a$ is the angular momentum per unit mass.
The signature in (\ref{eq.metric}) is $(+,-,-,-)$.
In (\ref{eq.tme}) $s$ denotes the spin weight of the field $\psi^{(s)}$,
$T^{(s)}$ is a source term, $\nabla_\mu$ denotes the Levi--Civita covariant derivation, 
$\Psi_2  =  -M/(r-\ii a \cos\theta)^3$ 
is the spin weight $0$ Weyl scalar of the Kerr metric in Kinnersley tetrad 
(see \cite{BCJR} for explicit expressions for the Kinnersley tetrad),
and $\Gamma^\mu$ is the ``connection vector''
\begin{eqnarray}
&& \Gamma^t= -\frac{1}{\Sigma}\left[\frac{M(r^2-a^2)}{\Delta}-(r+\ii a\cos\theta)   \right]\\
&& \Gamma^r = -\frac{1}{\Sigma}(r-M)\\
&& \Gamma^\theta = 0\\
&& \Gamma^\phi = -\frac{1}{\Sigma}\left[\frac{a(r-M)}{\Delta} +\ii\frac{\cos\theta}{\sin^2\theta}  \right]
\end{eqnarray}
introduced in \cite{BCJR}. The relation between $\psi^{(s)}$, $T^{(s)}$ and the Maxwell, linearized gravitational, neutrino and 
spin-$3/2$ fields and their sources is also explained in \cite{BCJR} and in further references cited there.

$g_{\mu\nu}$, $\Gamma^\mu$ and $\Psi_2$ are invariant under time translations and rotations generated by
the vector fields
$(\partial_t)^\mu$ and $(\partial_\phi)^\mu$, thus these time translations and rotations are symmetries of the 
TME if $T^{(s)}$ is also invariant under them. 
Furthermore, if $T^{(s)}=0$, then $\nabla_t$ and $\nabla_\phi$ are symmetry operators of the TME.
If $T^{(s)}$ is invariant under time translations or rotations, then $\nabla_t\psi^{(s)}$ or $\nabla_\phi\psi^{(s)}$
is a solution of the sourceless TME.
The TME has nontrivial second order symmetry operators as well (see Section 5.4.1 of \cite{Aphd}),
and there is a differential
operator that maps solutions of the spin-$s$ equation into solutions of the spin-$(-s)$ equation \cite{Wald}. 

The TME also has a discrete symmetry:
the transformation $P: \psi^{(s)}(t,r,\theta,\phi)\to \psi^{(s)}(t,r,\pi-\theta,\phi)^*$ 
is a discrete symmetry of the TME if 
$T^{(s)}$ is also invariant under this transformation, because
$g_{\mu\nu}$ is real and is invariant under the reflection $\theta\to \pi-\theta$ with respect to the equatorial plane, 
and $\Gamma^\mu$ and $\Psi_2$ are invariant under $\theta\to \pi-\theta$ followed by a complex conjugation.

Although the metric, $\Psi_2$, $\Gamma^\mu$ and $s$ take particular values in (\ref{eq.tme}), 
many of the following arguments are valid for arbitrary values of these quantities, 
restricted only by invariance requirements when necessary.

In the next section, we discuss the construction of the conserved currents that follow from the time translation and rotation symmetries of the TME in three partly different approaches. The current that follows from scaling symmetry is discussed in Section \ref{sec.teuk2}. 
Conserved currents related to the Sommerfeld boundary condition are constructed in Section \ref{sec.tmebc}.

\subsection{Energy- and angular momentum-like currents} 
\label{sec.teuk1}

For the application of the construction described in Section \ref{sec.sde}, 
let us multiply (\ref{eq.tme}) by $\sqrt{-g}$, where $g$ is the determinant of the metric. 
The Lagrangian function corresponding to the density (\ref{eq.l}) then takes the form
\beq
\label{eq.tlagr}
\fl \hspace{5mm}
\hat{\mc{L}}=\int \intd r\, \intd \theta\, \intd \phi \, \sqrt{-g}\,\bigl[
\psi^{(-s)}[(\nabla^\mu + s\Gamma^\mu)(\nabla_\mu +s\Gamma_\mu)-4s^2\Psi_2]\psi^{(s)}
-4\pi T^{(s)}\psi^{(-s)}\bigr]\, ,
\eeq
where $\psi^{(-s)}$ denotes the auxiliary field (this notation is justified below). $\hat{\mc{L}}$ can be converted into
\beq \hspace{5mm}
\fl
\hat{\mc{L}}^{(-)}=\int \intd r\, \intd \theta\, \intd \phi \, \sqrt{-g}\,
\bigl[\psi^{(s)}[(\nabla^\mu - s\Gamma^\mu)(\nabla_\mu -s\Gamma_\mu)-4s^2\Psi_2]\psi^{(-s)}
-4\pi T^{(s)}\psi^{(-s)}\bigr]\, ,
\eeq
by adding the total divergence terms 
$-\sqrt{-g}\,\nabla^\mu[\psi^{(-s)}(\nabla_\mu+s\Gamma_\mu)\psi^{(s)}]$ and 
$-\sqrt{-g}\,\nabla_\mu[\psi^{(s)}(-\nabla^\mu\\ +s\Gamma^\mu)\psi^{(-s)}]$
to the integrand, 
and this shows that the Euler--Lagrange equation for $\psi^{(s)}$
is 
\beq
\label{eq.tmeaux}
\sqrt{-g}\,[(\nabla^\mu - s\Gamma^\mu)(\nabla_\mu -s\Gamma_\mu)-4s^2\Psi_2]\psi^{(-s)}=0\, ,
\eeq
which is the TME with spin weight $-s$ and zero source. Thus we have found that the Euler--Lagrange equations for 
$\hat{\mc{L}}$ consist of the TME (\ref{eq.tme}) and another TME with opposite spin weight and zero source. 
This result also means that a pair of sourceless TMEs with opposite spin weight and multiplied by $\sqrt{-g}$ 
constitute a selfadjoint system of equations, 
which was observed in \cite{Wald} as well (see also \cite{AB}).

If $g_{\mu\nu}$, $\Psi_2$, $\Gamma^\mu$ and $T^{(s)}$ are invariant under the time translations and rotations generated 
by $(\partial_t)^\mu$ and $(\partial_\phi)^\mu$, which will also be denoted by $h^\mu$,
then the time translations and rotations, under which $\delta\psi^{(s)}$
is $-\partial_t \psi^{(s)}$ and $-\partial_\phi \psi^{(s)}$, are symmetries of (\ref{eq.tme}) (multiplied by $\sqrt{-g}$) 
according to the definition in Section \ref{sec.sde}, and the associated currents
\bea
\label{eq.e1}
\hat{\mc{E}}^\mu & = &  
-\psi^{(-s)}(\nabla^\mu+s\Gamma^\mu)\nabla_t\psi^{(s)}+\nabla_t\psi^{(s)}(\nabla^\mu-s\Gamma^\mu)\psi^{(-s)} \\
\label{eq.j1}
\hat{\mc{J}}^\mu & = &
-\psi^{(-s)}(\nabla^\mu+s\Gamma^\mu)\nabla_\phi\psi^{(s)}+\nabla_\phi\psi^{(s)}(\nabla^\mu-s\Gamma^\mu)\psi^{(-s)} 
\eea
are conserved if $\psi^{(s)}$ is a solution of the TME (\ref{eq.tme}) with spin weight $s$ and $\psi^{(-s)}$ is
also a solution of the TME with opposite spin weight and zero source.
We note that after applying (\ref{eq.elc2}), we divided the obtained currents by $\sqrt{-g}$, therefore the conservation equations for 
$\hat{\mc{E}}^\mu$ and $\hat{\mc{J}}^\mu$ are $\nabla_\mu\hat{\mc{E}}^\mu=0$ and $\nabla_\mu\hat{\mc{J}}^\mu=0$.
It is also important to note that it is not necessary to require any 
relation between $\psi^{(-s)}$ and $\psi^{(s)}$ for the conservation of $\hat{\mc{E}}^\mu$ and $\hat{\mc{J}}^\mu$.
The TME reduces to the Klein--Gordon equation in the case $s=0$, 
nevertheless $\psi^{(-s)}$ and $\psi^{(s)}$ are two independent fields even in this case.
$\hat{\mc{L}}$, $\hat{\mc{L}}^{(-)}$, $\hat{\mc{E}}^\mu$ and $\hat{\mc{J}}^\mu$ are complex, and since the real and imaginary parts of 
$\hat{\mc{E}}^\mu$ and $\hat{\mc{J}}^\mu$ are conserved separately, $\hat{\mc{E}}^\mu$ and $\hat{\mc{J}}^\mu$ comprise four real conserved currents.

Symmetry operators can be used to obtain further conserved currents;
if $O$ is a symmetry operator of the TME, then $\hat{\mc{E}}^\mu [O\psi^{(-s)}, \psi^{(s)}]$ and 
$\hat{\mc{J}}^\mu [O\psi^{(-s)}, \psi^{(s)}]$ are also conserved currents. 
If $T^{(s)}=0$ and $O_1$ and $O_2$ are symmetry operators of the TME, then
$\hat{\mc{E}}^\mu [O_1 \psi^{(-s)}, O_2\psi^{(s)}]$ and 
$\hat{\mc{J}}^\mu [O_1 \psi^{(-s)}, O_2 \psi^{(s)}]$ are conserved currents as well.
$O$, $O_1$ and $O_2$ can be any products of the symmetry operators mentioned before Section \ref{sec.teuk1}. 
Even if $T^{(s)}\ne 0$, if $T^{(s)}$ is invariant under time translations and rotations, then 
$\hat{\mc{E}}^\mu [O_1 \psi^{(-s)}, O_2\psi^{(s)}]$ and 
$\hat{\mc{J}}^\mu [O_1 \psi^{(-s)}, O_2 \psi^{(s)}]$ are conserved currents, where $O_2=\nabla_t^k \nabla_\phi^l$ and $k$ and $l$ are
arbitrary nonnegative integers.

Although (\ref{eq.l}) does not produce any source term in (\ref{eq.tmeaux}), the source $4\pi T^{(-s)}$ can be introduced into it by 
adding the term $-\sqrt{-g}\, 4\pi T^{(-s)}\psi^{(s)}$ to the Lagrangian density function. Furthermore, the Lagrangian  
can be brought to first order form by adding a total divergence. In this way one finds that   
\bea
\label{eq.tlagr2}
&& \mc{L}=\int \intd r\, \intd \theta\, \intd \phi \, 
\sqrt{-g}\, [-(\nabla_\mu-s\Gamma_\mu) \psi^{(-s)} (\nabla^\mu+s\Gamma^\mu) \psi^{(s)} 
- 4s^2 \Psi_2 \psi^{(-s)}\psi^{(s)} \nonumber \\ 
&& \hspace{4.5cm}-\,4\pi T^{(s)}\psi^{(-s)}-4\pi T^{(-s)}\psi^{(s)}]
\eea
is a Lagrangian for a pair of Teukolsky Master Equations with opposite spin weights. The source terms $T^{(s)}$ and 
$T^{(-s)}$ can be different even when $s=0$, and since $\psi^{(-s)}$ and $\psi^{(s)}$ are independent fields, (\ref{eq.tlagr2})
does not reduce to the usual Lagrangian of the scalar field at $s=0$.

Assuming that $T^{(-s)}$ is also invariant under time translations and rotations, one can apply the standard Noether construction
described in \ref{sec.nthr} to (\ref{eq.tlagr2}), 
with
$\delta\psi^{(\pm s)}=-h^\nu\partial_\nu \psi^{(\pm s)}$ and
$K^\mu=-h^\mu(\sqrt{-g}\cL)$, where 
\beq
\fl \hspace{3mm}
\cL = -(\nabla_\mu-s\Gamma_\mu) \psi^{(-s)} (\nabla^\mu+s\Gamma^\mu) \psi^{(s)} 
- 4s^2 \Psi_2 \psi^{(-s)}\psi^{(s)} 
-4\pi T^{(s)}\psi^{(-s)}-4\pi T^{(-s)}\psi^{(s)}
\eeq
is the integrand in (\ref{eq.tlagr2}) divided by $\sqrt{-g}$.
For the Noether currents one obtains
\begin{eqnarray}
\mc{E}^\mu & = &  (\nabla^\mu-s\Gamma^\mu)\psi^{(-s)} \nabla_t \psi^{(s)} + 
(\nabla^\mu+s\Gamma^\mu)\psi^{(s)} \nabla_t \psi^{(-s)} + (\partial_t)^\mu \cL  \nonumber\\
& = & {\mc{T}^\mu}_\nu (\partial_t)^\nu
\label{eq.E}
\end{eqnarray} 
\begin{eqnarray}
\mc{J}^\mu  & = &  (\nabla^\mu-s\Gamma^\mu)\psi^{(-s)} \nabla_\phi \psi^{(s)} + 
(\nabla^\mu+s\Gamma^\mu)\psi^{(s)} \nabla_\phi \psi^{(-s)} + (\partial_\phi)^\mu \cL
\nonumber \\
& = & {\mc{T}^\mu}_\nu (\partial_\phi)^\nu\ ,
\label{eq.J}
\end{eqnarray} 
where
\beq
\mc{T}^{\mu\nu}=  (\nabla^\mu-s\Gamma^\mu)\psi^{(-s)} \nabla^\nu \psi^{(s)} + 
(\nabla^\mu+s\Gamma^\mu)\psi^{(s)} \nabla^\nu \psi^{(-s)} + g^{\mu\nu} \cL\, . 
\eeq
It should be noted that this $\mc{T}^{\mu\nu}$ is not symmetric. If $T^{(-s)}=0$, 
then the integrands in $\mc{L}$ and $\hat{\mc{L}}$ differ in a total divergence only, therefore one expects that in this case the differences
$\mc{E}^\mu - \hat{\mc{E}}^\mu$ and $\mc{J}^\mu - \hat{\mc{J}}^\mu$ are identically conserved currents, 
i.e.\ currents of the form $\nabla_\nu \Sigma^{\mu\nu}$, where $\Sigma^{\mu\nu}$ is antisymmetric.
Indeed, it is not difficult to verify that the differences $\mc{E}^\mu - \hat{\mc{E}}^\mu$ and $\mc{J}^\mu - \hat{\mc{J}}^\mu$ are equal to $\nabla_\nu \Sigma^{\mu\nu}$ with 
$\Sigma^{\mu\nu}=   h^\nu \psi^{(-s)} (\nabla^\mu+s\Gamma^\mu)\psi^{(s)} -  h^\mu \psi^{(-s)} (\nabla^\nu+s\Gamma^\nu)\psi^{(s)}$
if $\psi^{(s)}$ satisfies (\ref{eq.tme}) and $\psi^{(-s)}$ satisfies the TME with spin weight $-s$ and $T^{(-s)}=0$.

If $T^{(-s)}$ or $T^{(s)}$ is zero, then further conserved currents can again be obtained by the replacements 
$\psi^{(-s)}\to O_1 \psi^{(-s)}$ or $\psi^{(s)}\to O_2 \psi^{(s)}$ in 
$\mc{E}^\mu [\psi^{(-s)}, \psi^{(s)}]$ and $\mc{J}^\mu [\psi^{(-s)}, \psi^{(s)}]$. 

The Lagrangian (\ref{eq.tlagr2}) also provides an opportunity to apply a further version of Noether's theorem, which is a generalization
of the usual construction of currents associated with spacetime symmetries in general relativity. 
In the usual construction, the conserved current associated with a Killing vector field $h^\mu$ is $T^{\mu\nu}h_\nu$,
where $T^{\mu\nu}$ is the energy-momentum tensor \cite{WaldGR,IW,Szabados}. However, this construction is not suitable for (\ref{eq.tlagr2}),
because the corresponding energy-momentum tensor $T^{\mu\nu}= \frac{-2}{\sqrt{-g}}\frac{\delta L}{\delta g_{\mu\nu}}$ is not divergenceless (i.e.\ $\nabla_\mu T^{\mu\nu}\ne 0$). The divergencelessness of $T^{\mu\nu}$ generally follows from the diffeomorphism symmetry of the Lagrangian, but (\ref{eq.tlagr2}) does not have complete diffeomorphism symmetry due to the presence of 
$\Psi_2$, $\Gamma^\mu$, $T^{(s)}$ and $T^{(-s)}$, which do not count as field variables. 
This can be remedied by taking also $\Psi_2$, $\Gamma^\mu$, $T^{(s)}$ and $T^{(-s)}$ to be field variables, 
but the divergencelessness of $T^{\mu\nu}$ is not guaranteed unless all fields except $g_{\mu\nu}$ satisfy their Euler--Lagrange equations, and the latter condition is violated by $\Psi_2$, $\Gamma^\mu$, $T^{(s)}$ and $T^{(-s)}$.
From here one can proceed by applying a generalization of the usual construction, 
which can be used when general kinds of fixed fields, not just $g_{\mu\nu}$, are present,
and which is described in detail in \cite{T} and also appears in more special form in the earlier papers \cite{Borokhov,DM1,DM2,NQV}.
This gives a current associated with $h^\mu$, which is conserved if $h^\mu$ is a Killing vector field and 
$\Psi_2$, $\Gamma^\mu$, $T^{(s)}$ and $T^{(-s)}$ are also invariant under the diffeomorphisms generated by $h^\mu$.
According to the generalized construction, the sought current is 
\beq
\label{eq.B}
\mc{B}^\mu = \frac{\delta L}{\delta \chi_j}\delta\chi_{j\nu}^\mu h^\nu\, ,
\eeq
where $L=\sqrt{-g}\,\cL$,
$\chi_j=\{ g_{\mu\nu},\Gamma_\mu,\Psi_2, T^{(s)}, T^{(-s)} \}$ denotes collectively the fixed fields 
(which are not required to satisfy their Euler--Lagrange equations), and
$\delta\chi_{j\nu}^\mu$ are quantities that appear in the transformation rules 
\beq
\delta\chi_j=\delta\chi_{j\nu}h^\nu + \delta\chi_{j\nu}^\mu \partial_\mu h^\nu
\eeq
of $\chi_j$ under diffeomorphisms. For convenience, we use $\Gamma_\mu$ instead of $\Gamma^\mu$ as an independent field variable,
but $\Gamma^\mu$ would be equally suitable.
The specific transformation rules are 
\beq
\delta g_{\lambda\rho} = -\nabla_\lambda h_\rho - \nabla_\rho h_\lambda\, ,\qquad 
\delta \Gamma_\lambda = -h^\nu \nabla_\nu \Gamma_\lambda - \nabla_\lambda h^\nu \Gamma_\nu\, ,
\eeq
\beq
\delta\Psi_2 = -h^\nu \partial_\nu\Psi_2\, ,\qquad \delta T^{(\pm s)}=-h^\nu\partial_\nu T^{(\pm s)}\, ,
\eeq
thus $\delta\Psi_{2\nu}^\mu = \delta T^{(\pm s)\mu}_{\nu} = 0$ and
\beq
\delta g_{\lambda\rho\nu}^\mu = -\delta_\lambda^\mu g_{\rho\nu}-\delta^\mu_\rho g_{\lambda\nu}\, ,\qquad
\delta \Gamma_{\lambda\nu}^\mu = -\Gamma_\nu \delta_\lambda^\mu\, ,
\eeq
and thus $\mc{B}^\mu$ takes the form
\beq
\mc{B}^\mu = -\sqrt{-g}\, \left(\frac{1}{2}T^{\lambda\rho}\delta g_{\lambda\rho\nu}^\mu +
J_{\Gamma}^\lambda \delta \Gamma_{\lambda\nu}^\mu \right) h^\nu
= \sqrt{-g}\, (T\indices{^\mu_\nu} + J_{\Gamma}^\mu \Gamma_\nu) h^\nu\, ,
\eeq
where
\bea
T^{\mu\nu} & = & \frac{-2}{\sqrt{-g}}\frac{\delta L}{\delta g_{\mu\nu}} \nonumber \\
& = &
-(\nabla^\mu-s\Gamma^\mu)\psi^{(-s)}(\nabla^\nu+s\Gamma^\nu)\psi^{(s)}
-(\nabla^\nu-s\Gamma^\nu)\psi^{(-s)}(\nabla^\mu+s\Gamma^\mu)\psi^{(s)}  \nonumber \\
&& -g^{\mu\nu}\cL
\label{eq.t}
\eea
\beq
J_{\Gamma}^\nu\ =\ \frac{-1}{\sqrt{-g}}\frac{\delta L}{\delta \Gamma_\nu} \ = \ 
 -s\psi^{(-s)}(\nabla^\nu +s\Gamma^\nu)\psi^{(s)}
+s\psi^{(s)}(\nabla^\nu-s\Gamma^\nu)\psi^{(-s)}\, .
\label{eq.jgamma}
\eeq
Comparing this result with $\mc{E}^\mu$ and $\mc{J}^\mu$, one sees that 
$\mc{B}^\mu=-\sqrt{-g}\,\mc{E}^\mu$ and $\mc{B}^\mu=-\sqrt{-g}\,\mc{J}^\mu$ for 
$h^\mu=(\partial_t)^\mu$ and $h^\mu=(\partial_\phi)^\mu$, i.e.\ the same currents are obtained as in the previous approach.
The relation between $\mc{T}^{\mu\nu}$ and $T^{\mu\nu}$ is $\mc{T}^{\mu\nu}=-(T^{\mu\nu}+J_\Gamma^\mu \Gamma^\nu)$.
We note that adding total divergences to $L$ does not destroy its diffeomorphism symmetry, 
and the right hand side of (\ref{eq.B}) depends on $L$ only through its Euler--Lagrange derivatives, 
therefore modifying $L$ by adding total divergences does not change $\mc{B}^\mu$.

\subsection{The conserved current associated with scaling transformations}
\label{sec.teuk2}

If the source term is zero in (\ref{eq.tme}), then the rescalings $\psi^{(s)}\to \ee^{\varsigma C}\psi^{(s)}$ are also symmetries of (\ref{eq.tme}) for any complex number $C$. The first order variation of $\psi^{(s)}$ is $\delta\psi^{(s)}=C\psi^{(s)}$
under these rescalings. The factor $C$ is not of much significance, therefore we set it to $1$.
The conserved current given by (\ref{eq.elc2}), after dividing by $\sqrt{-g}$, is then
\beq
\label{eq.S}
\hat{\mc{S}}^\mu = \psi^{(-s)}(\nabla^\mu+s\Gamma^\mu)\psi^{(s)} - \psi^{(s)}(\nabla^\mu-s\Gamma^\mu)\psi^{(-s)}\, .
\eeq
We note that in the special case of $s=1$, this current was also found in \cite{Aphd} (see Proposition 5.2.4.).

The first order Lagrangian $\mc{L}$ and the standard Noether construction can also be used to obtain the conserved current associated with rescalings.
If $T^{(\pm s)}=0$, then $L=\sqrt{-g}\,\cL$ satisfies the symmetry condition (\ref{eq.K}) with $\delta\psi^{(\pm s)}=\pm C\psi^{(\pm s)}$ and $K^\mu=0$ for any $\psi^{(\pm s)}$.
The Noether current given by (\ref{eq.n2}) and (\ref{eq.elc}) turns out to be identical with $\hat{\mc{S}}^\mu$.

As in Section \ref{sec.teuk1}, further conserved currents can be obtained by replacing 
$\psi^{(-s)}$ or $\psi^{(s)}$ with $O_1 \psi^{(-s)}$
and $O_2 \psi^{(s)}$ in 
$\hat{\mc{S}}^\mu [\psi^{(-s)}, \psi^{(s)}]$.  
It should be noted, however, that many of these currents are not new, because
\bea
\label{eq.se}
\hat{\mc{S}}^\mu [\psi^{(-s)}, \nabla_t\psi^{(s)}] & = & -\hat{\mc{E}}^\mu[\psi^{(-s)}, \psi^{(s)}] \\ 
\label{eq.sj}
\hat{\mc{S}}^\mu [\psi^{(-s)}, \nabla_\phi\psi^{(s)}] & = & -\hat{\mc{J}}^\mu[\psi^{(-s)}, \psi^{(s)}]\, ,
\eea
in accordance with the general considerations in the last paragraph before Section \ref{sec.bcurrents}.

If $T^{(s)}$ or $T^{(-s)}$ is not zero, then $\hat{\mc{S}}^\mu$ is not conserved, but
\beq
\label{eq.nc1}
\nabla_\mu \hat{\mc{S}}^\mu = 4\pi (\psi^{(-s)}T^{(s)} - \psi^{(s)}T^{(-s)})
\eeq
holds, if $\psi^{(\pm s)}$ satisfy the TME. 
Similarly, 
\bea
\label{eq.nc2}
\nabla_\mu\hat{\mc{E}}^\mu & = & -4\pi  (\psi^{(-s)}\nabla_t T^{(s)} - \nabla_t\psi^{(s)}T^{(-s)})\\
\label{eq.nc3}
\nabla_\mu\hat{\mc{J}}^\mu & = & -4\pi  (\psi^{(-s)}\nabla_\phi T^{(s)} - \nabla_\phi\psi^{(s)}T^{(-s)})
\eea
and
\bea
\label{eq.nc4}
\nabla_\mu\mc{E}^\mu & = & -4\pi  (\psi^{(-s)}\nabla_t T^{(s)} + \psi^{(s)}\nabla_t T^{(-s)})\\
\label{eq.nc5}
\nabla_\mu\mc{J}^\mu & = & -4\pi  (\psi^{(-s)}\nabla_\phi T^{(s)} + \psi^{(s)}\nabla_\phi T^{(-s)})
\eea
hold for general $T^{(\pm s)}$.
(\ref{eq.nc1})-(\ref{eq.nc5}) can be derived either by direct calculation or with the help of \ref{sec.nthr}.
These identities are still useful for testing numerical solutions of the TME. They can also be converted to charge balance equations,
which contain additional terms, corresponding to the right hand sides of (\ref{eq.nc1})-(\ref{eq.nc5}), 
that give the amount by which the conservation of the relevant charges is violated.

\subsection{Conserved currents associated with the Sommerfeld boundary condition}
\label{sec.tmebc}

The Sommerfeld boundary condition 
takes a simple form if the tortoise coordinate $r^*$, defined as $dr^*=\frac{r^2+a^2}{\Delta}dr$, is used instead of $r$. 
At an outer boundary $\mc{B}$ it reads
\beq
\label{eq.sbc}
(\partial_t\psi^{(s)}(t,r^*,\theta,\phi)+\partial_{r^*}\psi^{(s)}(t,r^*,\theta,\phi))|_{\mc{B}}=0\, ,
\eeq
whereas at an inner boundary there is a $-$ sign in front of the second term.
(\ref{eq.sbc}) does not follow from a Lagrangian in the sense of \ref{sec.bcurr}, but the construction described in Section \ref{sec.bcurrents} can be applied to it. The auxiliary Lagrangian for (\ref{eq.sbc}) is 
$L_{\mc{B}}= (\partial_t\psi^{(s)}(t,r^*,\theta,\phi)+\partial_{r^*}\psi^{(s)}(t,r^*,\theta,\phi)) \rho_\mc{B}$.
The boundary Euler--Lagrange equation $\frac{\delta L_{\mc{B}}}{\delta \psi^{(s)}}|_{\mc{B}}=0$ is thus
\beq
\label{eq.sbcrho}
-(\partial_t\rho_{\mc{B}} +\partial_{r^*}\rho_\mc{B})|_{\mc{B}} =0\, , 
\eeq
i.e.\ $\rho_\mc{B}$ is required to satisfy the same equation as $\psi^{(s)}$
at the boundary. Apart from (\ref{eq.sbcrho}), there is no restriction on $\rho_\mc{B}$.
For $j_{\mc{B}}^\mu$ one obtains the expression $(\delta_t^\mu+\delta_{r^*}^\mu)\rho_{\mc{B}}\delta\psi^{(s)}$.
(\ref{eq.sbc}) has scaling symmetry, under which $\delta\psi^{(s)} = \psi^{(s)}$, thus the corresponding conserved current is
\beq
\label{eq.bS}
\mc{S}^\mu_{\mc{B}}[\rho_{\mc{B}},\psi^{(s)}]=(\delta_t^\mu+\delta_{r^*}^\mu)\rho_{\mc{B}}\psi^{(s)}\, .
\eeq
The conservation law for $\mc{S}^\mu_{\mc{B}}$ is $(\partial_\mu \mc{S}^\mu_{\mc{B}})|_{\mc{B}}=0$.
If $\psi^{(s)}$ satisfies (\ref{eq.sbc}), then its first and higher partial derivatives with respect to the coordinates also satisfy 
(\ref{eq.sbc}), therefore one can obtain further conserved boundary currents by replacing $\psi^{(s)}$ with these derivatives in 
$\mc{S}^\mu_{\mc{B}}[\rho_{\mc{B}},\psi^{(s)}]$. It should be noted that 
since $\rho_{\mc{B}}$ and $\psi^{(s)}$ satisfy the same boundary conditions, it is possible to choose $\rho_{\mc{B}} = \psi^{(s)}$.
For an inner boundary there is an obvious sign change in (\ref{eq.bS}), i.e.\ 
$\mc{S}^\mu_{\mc{B}}[\rho_{\mc{B}},\psi^{(s)}]=(\delta_t^\mu-\delta_{r^*}^\mu)\rho_{\mc{B}}\psi^{(s)}$.

\section{Concluding remarks}

In this paper conserved currents for the TME corresponding to its time translation, rotation and scaling symmetries (see (\ref{eq.e1}), (\ref{eq.j1}), (\ref{eq.E}), (\ref{eq.J}) and (\ref{eq.S})) were constructed.
From these primary currents, which involve two independent solutions of the TME with opposite spin weights,
infinitely many further conserved currents can also be obtained with the help of symmetry operators.

The main potential application that we had in mind for the conserved currents is the verification of numerical solutions of the TME.
Verifying numerical solutions of differential equations is one of the usual applications of conserved currents,
and in the case of the Klein--Gordon equation in Kerr spacetime energy and angular momentum currents have already been used 
for this purpose. 
However, an important difference between the Klein--Gordon equation and the TME is that the latter 
does not follow from a Lagrangian, which is an obstacle to finding conserved currents. 
In order to overcome this difficulty we applied a variant of Noether's theorem, 
which is not restricted to differential equations that follow from a Lagrangian. Although this variant has been known for some time, 
it is considerably less well known than the usual version of Noether's theorem. 
For this reason we briefly reviewed it in general form,
thus in addition to dealing with the particular case of the TME we also indicated 
how other differential equations and symmetries could be dealt with. 

A further difference between the Klein--Gordon equation and the TME is that the latter is a complex equation, 
i.e.\ its complex conjugate is different from itself. The conserved currents found for the TME are also complex 
and do not involve a complex conjugation, 
in contrast with the energy and angular momentum currents of a complex Klein--Gordon field. 
The TME also contains a source term, which has to be zero or satisfy appropriate symmetry conditions in order
that (\ref{eq.e1}), (\ref{eq.j1}), (\ref{eq.E}), (\ref{eq.J}) and (\ref{eq.S}) be conserved. For general source terms
(\ref{eq.nc1})-(\ref{eq.nc5}) hold instead of the usual current conservation equations, but these identities are still 
useful for the purpose of testing numerical computations.

A physical system may include boundary conditions for the fields that it contains, 
and imposing boundary conditions may also be necessary in numerical calculations 
because of the finiteness of the computational domain. 
We extended Noether's standard theorem and its variant mentioned above to boundary conditions and their symmetries, 
providing a possibility to use symmetries and conserved currents for testing whether the boundary conditions are also satisfied
by a numerical solution. Nevertheless, such tests can be expected to be useful mainly for complicated boundary conditions,
whereas for the usual relatively simple boundary conditions they are less important. 
Although we concentrated on boundary conditions, conditions at interfaces could be handled in a similar way.

A further interesting problem is to determine if the currents found in the present paper are useful for obtaining decay estimates
for the Maxwell and linearized gravitational fields. $\psi^{(s)}$ and $\psi^{(-s)}$ would not be independent in such applications,
and it is likely that their relation would also have to be taken into account. It would also be interesting to investigate whether 
there are further conserved currents of the TME beyond those mentioned in this paper.

\ack

I would like to thank Istv\'an R\'acz, Lars Andersson, Andr\'as L\'aszl\'o and K\'aroly Csuk\'as for useful discussions on the Teukolsky Master Equation. I thank one of the referees for drawing my attention to boundary conditions.
I acknowledge support by the NKFIH grant no.\ K116505.

\appendix

\section{The Noether construction} 
\label{sec.nthr}

In this appendix, the standard construction of conserved currents associated with continuous 
Lagrangian symmetries is recalled in a 
modern and general form, allowing Lagrangians that depend on arbitrarily high derivatives of the fields, general kinds of symmetry transformations, and anticommuting (Grassmann algebra valued) fields. 
References where further details can be found are \cite{Olver,Noether,Y,BCA}, for example.
In addition, conserved currents associated with boundary conditions and their symmetries are introduced in \ref{sec.bcurr}.
Some of the notation used below is introduced at the beginning of Section \ref{sec.sde}. 

Let us consider an action
\beq
S=\int_U d^{D+1}x\, L(x^\mu,\Phi_i(x^\mu),\partial_\nu\Phi_i(x^\mu),\partial_{\nu\lambda}\Phi_i(x^\mu),\dots)\, ,
\eeq
where $d^{D+1}x$ is the measure determined by the coordinates $x^\mu$
and $L$ is the Lagrangian density function, which is allowed to depend on arbitrarily high derivatives of the fields. 
$L$ is assumed to be even, regarding commutation properties. 
The Lagrangian function $\int dx^1\dots dx^D\, L$ is denoted by $\mc{L}$.
If $\Phi_i$ is complex for some value of $i$, then $L$ should generally be allowed to depend also on $\Phi_i^*$ and its derivatives.
In order to keep the formulas shorter, we omit $\Phi_i^*$ and its derivatives, but it would be straightforward include them.
Complex fields can also be taken into account as two real fields.

Next, let us consider a one-parameter family of transformations of the fields. They may form a one-parameter transformation group, 
but this is not required.
After linearization in the parameter, denoted by $\varsigma$, 
the transformations can be written as
\beq
\label{eq.tr1}
\Phi_i\to \Phi_i+\varsigma\,\delta \Phi_i\ .
\eeq
$\varsigma$ is assumed to be real number valued and $\delta \Phi_i$ is assumed to have the same commutation character 
as $\Phi_i$.
Usually $\delta \Phi_i$ is a local function of the fields.  
A field configuration\footnote{By field configuration we mean all values of the fields in an open domain in $M$, not just on a hypersurface.} is said to be invariant under the transformation (\ref{eq.tr1}) if $\delta \Phi_i=0$ 
holds for the configuration. 
(\ref{eq.tr1}) is induced in many important cases by transformations in the base manifold or in the target space of the fields, 
but it may be more general.  
The associated first order variation of $L$ is defined as
$ \delta L = \frac{dL[\Phi+\varsigma\,\delta\Phi]}{d\varsigma}|_{\varsigma=0}$, and
\beq
\delta L  =  \frac{\partial L}{\partial \Phi_i}\delta\Phi_i +
\frac{\partial L}{\partial (\partial_\mu\Phi_i)}\partial_\mu\delta \Phi_i +
\frac{\partial L}{\partial (\partial_{\mu\nu}\Phi_i)}\partial_{\mu\nu} \delta \Phi_i + \dots\, . 
\label{eq.dl1} 
\eeq
$\delta L$ can be rewritten as
\beq
\label{eq.dl3}
\delta L[\Phi,\delta\Phi] =  \mb{E}[\Phi]^i\delta\Phi_i 
+ \partial_\mu j^\mu[\Phi,\delta\Phi]\ , 
\eeq
where 
\beq
\label{eq.el1}
\mb{E}[\Phi]^i=
\frac{\delta L}{\delta \Phi_i} =
\frac{\partial L}{\partial \Phi_i}-\partial_\mu\frac{\partial L}{\partial (\partial_\mu\Phi_i)}
+\partial_{\mu\nu}\frac{\partial L}{\partial (\partial_{\mu\nu}\Phi_i)}-
\partial_{\mu\nu\lambda}\frac{\partial L}{\partial (\partial_{\mu\nu\lambda}\Phi_i)}+
\dots\, ,
\eeq
which is the Euler--Lagrange derivative of  $L$ with respect to $\Phi_i$, 
and
\bea
&& \hspace{-2cm} j^\mu[\Phi,\delta\Phi]  =   \frac{\partial L}{\partial (\partial_\mu\Phi_i)}\delta \Phi_i+
\Biggl(\frac{\partial L}{\partial (\partial_{\mu\nu}\Phi_i)}\partial_\nu \delta \Phi_i
-\partial_\nu\frac{\partial L}{\partial (\partial_{\mu\nu}\Phi_i)} \delta \Phi_i\Biggr)\nonumber\\
&& + \Biggl(\frac{\partial L}{\partial (\partial_{\mu\nu\lambda}\Phi_i)}\partial_{\nu\lambda}\delta\Phi_i
-\partial_\nu\frac{\partial L}{\partial (\partial_{\mu\nu\lambda}\Phi_i)}\partial_\lambda\delta\Phi_i
+\partial_{\nu\lambda}\frac{\partial L}{\partial (\partial_{\mu\nu\lambda}\Phi_i)}\delta\Phi_i\Biggr)+\dots\, . \nonumber \\
\label{eq.elc}
\eea
If
\beq
\label{eq.K}
\delta L= \partial_\mu K^\mu 
\eeq
holds for a configuration of the fields with some $K^\mu$, which is usually a local function of $\Phi_i$,
then 
(\ref{eq.dl3}) implies that
\beq
\label{eq.n1}
\partial_\mu J^\mu + \mb{E}^i \delta \Phi_i
= 0\ ,
\eeq
where $J^\mu$ is defined as
\beq
\label{eq.n2}
J^\mu = j^\mu -K^\mu
\eeq
and is called the Noether current associated with (\ref{eq.tr1}).
In particular, if $\Phi_i$ also satisfy their Euler--Lagrange equations, 
i.e.\ $\mb{E}[\Phi]^i=0$,
then from (\ref{eq.n1}) it follows that
the current $J^\mu$ is conserved: $\partial_\mu J^\mu=0$. 
Such a conservation law can be converted into a charge conservation law or balance equation using Stokes' theorem.
For testing numerical results both the charge conservation laws and the current conservation laws are suitable.

It is very important to note that although (\ref{eq.K}) is often assumed to be an identity that 
holds for any field configuration, this is not necessary and we do not require it in this paper. 
(\ref{eq.K}) may be an equality that holds only for $\Phi_i$ that satisfy the 
Euler--Lagrange equations, or only for an even more special class of configurations of $\Phi_i$.
The conservation of $J^\mu$ is stated, of course, only for those solutions of the Euler--Lagrange equations
that satisfy (\ref{eq.K}).
If (\ref{eq.K}) holds for all solutions of the Euler--Lagrange equations, 
then (\ref{eq.tr1}) is called a Lagrangian symmetry transformation. 
If (\ref{eq.K}) holds only for a subset of the solutions of the Euler--Lagrange equations, then 
one might call (\ref{eq.tr1}) a partial Lagrangian symmetry.

It is clear that $K^\mu$ is not uniquely determined in (\ref{eq.K}), therefore in the applications a reasonable choice should be made to fix $K^\mu$. There are many important cases in which it is possible to choose $K^\mu=0$. In Section \ref{sec.sde}, for example,
it is natural to choose $K^\mu=0$.

\subsection{Extension of Noether's theorem to boundary conditions}
\label{sec.bcurr}

Let us consider now the situation when the base manifold $M$ has some boundaries and some conditions are imposed on 
$\Phi_i$ at these boundaries. A boundary condition at a boundary $\mc{B}$ is a system of equations
that contains $\Phi_i$ and their derivatives at $\mc{B}$. The boundaries that we consider can have any dimension lower than $D+1$.
For example, if $M=\RR\times [-1,1]^3 $, where $[-1,1]$ is the closed interval with endpoints $-1,1$ (and thus $[-1,1]^3$ is a cube), 
then $M$ has boundary pieces of dimension $3$, $2$ and $1$.
We assume that $\Phi_i$ and sufficiently many of their derivatives can be extended continuously to $\mc{B}$ from
the interior of $M$, 
thus $\Phi_i$ and their derivatives at $\mc{B}$ can be obtained from their values in the interior of $M$. 
It is also always assumed that the domain on which the coordinates $x^\mu$ are defined contains the piece of $\mc{B}$ 
that is under consideration.

We say that a boundary condition at a boundary $\mc{B}$ follows from a Lagrangian if it takes the form
\beq 
\label{eq.bc}
\frac{\delta L_{\mc{B}}}{\delta \Phi_i}|_{\mc{B}} = 0\, ,
\eeq 
where $L_{\mc{B}}[\Phi]$ is a local function of the fields. 
$L_{\mc{B}}[\Phi]$ is assumed to be defined in a neighbourhood of $\mc{B}$, 
and it is allowed to depend on the derivatives of $\Phi_i$
along all directions, not only along $\mc{B}$. We call (\ref{eq.bc}) boundary Euler--Lagrange equations.

Symmetries of $L_{\mc{B}}[\Phi]$ can be defined similarly as symmetries of usual Lagrangians;
a transformation $\Phi_i\to \Phi_i+\varsigma\,\delta \Phi_i$ of the fields is a symmetry of $L_{\mc{B}}[\Phi]$
if 
\beq
\label{eq.Kb}
(\delta L_{\mc{B}} - \partial_\mu K_{\mc{B}}^\mu )|_{\mc{B}} = 0
\eeq
holds, with a suitable $K_{\mc{B}}^\mu$, for any solution of (\ref{eq.bc}) .
$\delta\Phi_i$, which characterize the symmetry transformation, and $K_{\mc{B}}^\mu$ are assumed to be 
defined in a neighbourhood of $\mc{B}$, and usually they are local functions of $\Phi_i$.
$j^\mu_{\mc{B}}$ is defined in the same way as $j^\mu$ (see \ref{eq.elc}), and the Noether current associated with $L_{\mc{B}}$
and the symmetry transformation $\Phi_i\to \Phi_i+\varsigma\,\delta \Phi_i$ is 
\beq
\label{eq.n2b}
J_{\mc{B}}^\mu = j_{\mc{B}}^\mu -K_{\mc{B}}^\mu\ .
\eeq
The conservation law for $J_{\mc{B}}^\mu$ is $(\partial_\mu J_{\mc{B}}^\mu)|_{\mc{B}}=0$, which holds
for any field configuration that satisfies the boundary condition (\ref{eq.bc}) and the symmetry condition (\ref{eq.Kb}).
It should be noted that $J_{\mc{B}}^\mu$ has $D+1$ components, and 
it is not necessarily tangential to $\mc{B}$.  
If $J_{\mc{B}}^\mu$ is tangential to $\mc{B}$, then the conservation of $J_{\mc{B}}^\mu$ implies a 
charge conservation law at $\mc{B}$ in virtue of Stokes' theorem. If $J_{\mc{B}}^\mu$ is not tangential to $\mc{B}$,
then one can still obtain a balance equation, but it contains an additional term, 
which gives the amount by which the conservation of the relevant charge is violated. 
In numerical calculations $(\partial_\mu J_{\mc{B}}^\mu)|_{\mc{B}}=0$ can be verified by calculating $J_{\mc{B}}^\mu$
and then $\partial_\mu J_{\mc{B}}^\mu$ in a neighbourhood of $\mc{B}$, and then checking if $\partial_\mu J_{\mc{B}}^\mu$
goes to zero at $\mc{B}$. 

Although conserved currents associated with boundary conditions that follow from a Lagrangian exist in principle, 
they may be almost trivial or of little practical use if the boundary condition is simple.
For example, the Dirichlet boundary condition $\Phi|_{x^1=0}=0$ at $x^1=0$ for a scalar field 
follows from the Lagrangian $L_{\mc{B}}=\frac{1}{2}\Phi^2$, which has translation symmetry in any direction $h^\mu$ 
(where $h^\mu$ is a constant vector) with $K^\mu_{\mc{B}}=-\frac{1}{2}h^\mu \Phi^2$, but the corresponding Noether currents $J_{\mc{B}}^\mu = \frac{1}{2} h^\mu \Phi^2$
are zero at $x^1=0$ if $\Phi|_{x^1=0}=0$. 
Nevertheless, the conservation laws $(\partial_\mu J_{\mc{B}}^\mu)|_{x^1=0} = (\frac{1}{2}h^\mu \partial_\mu \Phi^2)|_{x^1=0} =0$ are suitable, in principle, for checking whether $\Phi|_{x^1=0}=0$ holds. 
On the other hand, in practice it is not necessary to verify $\Phi|_{x^1=0}=0$ in this way.
We note that with $K^\mu_{\mc{B}}=-\frac{1}{2}h^\mu \Phi^2$ (\ref{eq.Kb}) holds for arbitrary field configurations.

There are also common boundary conditions that do not follow from a Lagrangian in the above sense; 
the Neumann boundary condition $\partial_1\Phi|_{x^1=0}=0$, for example, is one of them.
These cases are discussed further in Section \ref{sec.bcurrents}.

It should be noted that boundary conditions are often derived by applying a variational principle  
to an action defined on the base manifold including its boundaries. Both the equations of motion and the boundary conditions 
follow from a single action in this approach.
However, this setting did not appear useful to us
from the point of view of conserved currents associated with boundary conditions, 
therefore we have taken a different approach, 
in which the action principle is not used.


\end{document}